\documentclass[A4,11pt]{article}
\usepackage{graphicx}
\begin{document}

\title{Emergent Universe with Interacting fluids and Generalized Second Law of Thermodynamics}

\author{B. C. Paul\thanks{Electronic mail : bcpaul@iucaa.ernet.in}  \\
    Physics Department, North Bengal University, \\
Siliguri, Dist. : Darjeeling, Pin : 734 013, West Bengal, India \\
A. S. Majumdar\thanks{ Electronic mail : archan@bose.res.in}\\
S. N. Bose National Centre for Basic Sciences \\
Block JD, Sector -III, Salt Lake,
Kolkata-700098}

\date{}

\maketitle									

\vspace{0.5in}

\begin{abstract}
We investigate the emergent universe scenario in the presence of interacting 
fluids. The non-linear equation of state (EoS) considered in the general 
theory of relativity for obtaining emergent universe is effectively a 
cosmological model  with a composition of three fluids. In this paper we 
consider two models to realize viable cosmological scenarios, {\it viz.},
(i) a two-fluid model with interaction of a pressureless  fluid with the fluid 
having the non-linear EoS needed for the emergent universe,  and (ii) a 
three-fluid model with interaction among  the three fluids which originate 
from the EoS of the emergent universe. It is found that realistic cosmological 
models in accordance with observations are not ruled out for both the above 
cases. We further show that the generalized second law of thermodynamics is 
found to hold good in the emergent universe with interacting fluids.
\end{abstract}

\vspace{0.2cm}

PACS number(s) : 04.20.Jb, 98.80.Cq, 98.80.-k

\vspace{4.5cm}
\pagebreak

\section{ Introduction}

The astronomical and cosmological observations predict that  we live in an 
expanding Universe. After the discovery of the cosmic microwave background 
radiation (CMBR) \cite{penz}, the big-bang cosmology  has become the standard 
model for cosmology which has a beginning of the Universe at some finite past. 
It is found that the big-bang cosmology does face some problems in addressing 
issues of the observed universe both in the early and late universe. A number 
of problems,  namely, horizon problem, flatness problem, singularity 
problem etc. crop up when one probes the early universe in the framework of the
big-bang model. These  problems however can be resolved by evoking a phase 
of inflation \cite{guth,sato,linde,stein,al,linde1,linde2,mukh} at a very early
 epoch. Furthermore, it is known that the  large scale structure formation of the universe can be addressed in
 this scenario successfully. On the other hand, very recent observations 
predict that our universe is passing through a phase of acceleration 
\cite{riess, perl1,perl}. This phase of acceleration is believed to be a late 
time phase of the universe and may be accommodated in the standard model by 
adding a positive cosmological constant in the Einstein's field equation.
 
In spite of  its overwhelming success, modern big-bang cosmology still has 
some unresolved issues. The physics of the inflation and the introduction of 
a small cosmological constant for late time acceleration are  not completely 
understood \cite{sean,dicke} in  details. Moreover, various competing
models exist that are as yet not fully distinguished empirically from the
currently available observational data. This is why there is enough motivation 
to search for an alternative cosmological model.  In this context, Ellis and Maartens 
\cite{ellis} considered the possibility of a cosmological model \cite{har} in 
which there is no big-bang singularity,  no beginning of time, and the universe 
effectively get rid of a quantum regime for space-time by staying large at all 
times. The universe started out in the infinite past in an almost static 
Einstein universe, and subsequently, it entered in an  expanding phase slowly, 
eventually evolving into a hot big-bang era. Later Ellis, Murugan and 
Tsagas \cite{ ellis1} constructed an emergent universe scenario for a closed 
universe with a minimally coupled scalar field $\phi$, which has  a special 
form for the interaction potential  $V(\phi)$. It was pointed out later 
\cite{ellis2} that the potential is  similar to what one obtains from a 
modified gravitational action with a polynomial Lagrangian $L= R + \alpha R^2$ 
after a suitable conformal transformation and identifying the field as $\phi= - \sqrt{3} \; \ln (1+ 2 \alpha R) $ with a negative $\alpha$.  

The EU scenario 
merits attention, as it promises to solve several conceptual as well as technical 
issues of the big-bang model. A notable direction is regarding the cosmological
constant problem \cite{padmas}.   Mukherjee {\it et al.} \cite{euorg} obtained an emergent universe in the framework of general theory of  relativity   in a flat universe with a nonlinear equation of state of the form 
\begin{equation}
 p = A \rho- B \sqrt{\rho}
\end{equation}
where $A$ and $B$ are arbitrary constants.  Such potentials have been also studied in the context of modified Chaplygin gas models \cite{chimento}. A spatially flat universe is most likely as predicted from recent cosmological  and astronomical observations.  Such  an emergent universe scenario can be 
realized also in a modified theory of gravitation \cite{eust} by including 
a Gauss-Bonnet term in the presence of dilaton coupling \cite{eugb}, Brane 
world gravity \cite{b1,b2,deb}, Brans-Dicke theory \cite{campo}, the
non-linear sigma model \cite{bee}, Chiral cosmological fields in Einstein Gauss-Bonnet gravity \cite{b1}, dark sector fields in a chiral cosmological model \cite{b2} and exact global phantonical solution \cite{b3}. Emergent universe accommodates a late time 
de-Sitter phase and thus it naturally leads to the late time acceleration of
the universe, as well. Such a scenario is promising from the perspective of
offering unified early as well as late time dynamics of the universe \cite{varun}. Note,
however, that the focal point of unification in such emergent universe models
lies in the choice of the equation of state for the polytropic fluid, while 
several other models of unification rely more on the scalar field dynamics
through choice of field potentials \cite{asm1,asm2,asm3}.

The emergent model proposed by Mukherjee {\it et al. } \cite{euorg} in which a polytropic equation of state (henceforth, EoS) is used,  gives rise
to a universe with a composition of three different types of  
fluids determined by the parameters $A$ and $B$.  In the original emergent universe model proposed by Mukherjee  {\it et al.} \cite{euorg}, it was assumed non-interacting fluids and each of the three types of fluids identified  satisfy  conservation equations separately.   Recently using 
the observational prediction of  WMAP7 \cite{bc1} and Planck2013  \cite{bc2} 
the permitted range of values of the parameters $A$ and $B$ are
determined 
\cite{pt1,pt2,pt3}. 
The recent cosmological observations  from Planck2013 impose tight bounds 
on the EoS parameters in an   emergent universe (EU) \cite{bc2}. For a viable 
cosmological scenario, 
it is further necessary to consider a consistent model of the universe which  
contains
 radiation dominated, matter dominated  and subsequently the late accelerated 
phases of the universe.  In the original EU model it is shown \cite{euorg} that   the composition of the universe is fixed once $A$ is fixed. 
A problem thus arises as to how a pressureless matter component could be
accommodated within such a scenario. However, allowing interaction among the constituent fluids of the emergent universe may open up richer
physical consequences.  

Interacting fluid cosmological models have been  
previously considered in the literature, and among a variety of reasons and motivations
for such models, analyses of interactions within the dark sector are quite popular  \cite{bar,cim,jam,lip,cost,cot}. In the present context
interaction among the constituent fluids is useful to obtain a consistent 
evolutionary scenario of the universe. Another important consistency condition
is imposed through the thermodynamics of an expanding universe. There
has been a lot of recent interest in the study of the connection between
thermodynamics and gravitational dynamics in the presence of horizons  
\cite{fro,fro1,fro2,fro3,fro4,gwg,jac,jac1,jac2}. Another aspect the Hawking temperature of the apparent horizon in a FRW universe was determined by Cai {\it et al.} \cite{cai}.  Indeed, Einstein's equations
have been interpreted as a thermodynamical relation resulting from the
displacement of the horizon. Since the emergent universe
scenario entails a phase of accelerated expansion, it is relevant here
to study the status of the second law of thermodynamics in the picture
involving interacting fluids in the emergent universe. 

With the above motivations, in the present
paper we   consider two different cases in the emergent universe scenario: 
(i) a two-fluid model with interaction of the fluid having the non-linear EoS given by eq. (1) with 
another barotropic  fluid, beginning  at  some time $t=t_i$ (Model-I) and (ii) a three-fluid model with interaction among the various constituent
fluids with different individual equations of state, starting at a time  $t= t_o$ (Model-II).
The paper is organized as follows : In Sec. 2 we set up the field equation for emergent universe in the general theory of relativity, and obtain cosmological  
solutions. In Sec. 3 we consider the above two cases, Model-I and Model-II.
We show that generation of pressureless matter fluids is possible within
the emergent universe scenario.
In Sec. 4, we demonstrate the consistency of the emergent universe scenario
with interacting fluids with the generalized second law of thermodynamics. 
Finally, we make some concluding remarks in Sec. 5.

\section {Field Equation and Cosmological Solution}

The Einstein field equation is given by
\begin{equation}
R_{\mu \nu} - \frac{1}{2} g_{\mu \nu} R = 8 \pi G T_{\mu \nu}
\label{a}
\end{equation}
where $R_{\mu \nu } $, $R$,  $g_{\mu \nu}$, $T_{\mu\nu}$  and $G$ represent the Ricci tensor, Ricci scalar, metric tensor, matter-energy tensor and Newton's gravitational constant.
Here we consider four dimensions for which $\mu, \; \nu$ runs from (0, 1, 2, 3). 

We consider a flat Robertson-Walker metric which is given by 
\begin{equation}
ds^2= -dt^2+ a^2(t)\left[ dr^2+ r^2 \left( d\theta^2 + sin^2 \theta d\phi^2 \right) \right]
\label{a1}
\end{equation}
where  $a (t)$ represents the scale factor of the Universe. We consider the energy momentum tensor as $T^{\mu}_{ \nu} = {\it diagonal} (\rho, -p, -p ,-p)$, where $\rho$ is the energy density and $p$ is the pressure.  
Using the flat Robertson-Walker metric given by eq. (3) in the Einstein's  field equation one obtains
\begin{equation}
\label{fr1}
\rho= 3 \left( \frac{\dot{a}}{a}\right)^{2},
\end{equation}
\begin{equation}
\label{fr1}
p =- \left[  2 \frac{\ddot{a}}{a} + \left( \frac{\dot{a}}{a}\right)^{2} \right]
\end{equation}
where we set $G= \frac{1}{8 \pi}$  and $c=1$. The  conservation equation is given by
\begin{equation}
\label{cvs}
\frac{d\rho}{dt}+3 H \left( p+\rho \right) = 0
\end{equation} 
where $H= \frac{\dot{a}}{a}$ represents the Hubble parameter. As mentioned earlier,  Mukherjee {\it et al.}  \cite{euorg} obtained an emergent universe  scenario with a polytropic equation of state (henceforth, EoS) given by
\begin{equation}
\label{eos1}
p= A \rho - B \rho^{1/2}
\end{equation}
where $A$ and $B$ are arbitrary constants. Making use of the conservation equation and the EoS given by eqs. (4) and (5) in eq. (7),  one obtains a second order differential equation given by
\begin{equation}
2 \frac{\ddot{a}}{a} + (3 A + 1) \left( \frac{\dot{a}}{a}\right)^{2} - \sqrt{3} B  \frac{\dot{a}}{a} =0.
\end{equation}
The scale factor of the universe is thus obtained integrating eq. (8)  which is given by
\begin{equation}
a(t) = \left[ \frac{3 K(A+1)}{2} \left( \sigma + \frac{2}{\sqrt{3}B } \; e^{\frac{\sqrt{3}}{2} B t} \right) \right]^{\frac{2}{3(A+1)}}
\end{equation}
where $K$ and $\sigma$ are the two integration constants. It is interesting to note that $B < 0$ leads to a contracting universe whereas with  $B > 0$ and   $A > -1$ leads to a non-singular solution which is expanding. The later solution corresponds to an emergent universe which was obtained by Mukherjee {\it et. al.} \cite{euorg}. 
The energy density of the universe in terms of scale factor is obtained from eq. (6) making use of EoS given by eq. (7) which is given by
\begin{equation}
\rho(a) = \frac{1}{(A+1)^2} \left(  B+ \frac{K}{a^{\frac{3(A+1)}{2}}} \right)^{2}.
\end{equation}
Expanding the above expression,  one obtains energy density as the sum of three terms which can be identified with three different types of fluids. Thus, the components of energy density and pressure can be expressed  as follows:
\begin{equation}
\rho(a) =  \Sigma_{i=1}^{3} \rho_i  \;  \;  \; and  \; \; \; p(a) =  \Sigma_{i=1}^{3} p_i
\end{equation}
where we denote
\begin{equation}
\label{r1}
\rho_1 = \frac{B^2}{(A+1)^2},  \; \;
\rho_2 = \frac{2 K B}{(A+1)^2}  \frac{1}{a^{\frac{3(A+1)}{2}}} ,  \; \; 
\rho_3 = \frac{K^2}{(A+1)^2} \frac{1}{a^{3(A+1)}} 
\end{equation}
\begin{equation}
\label{r2}
p_1=  = - \frac{B^2}{(A+1)^2},  \; \;
p_2 = \frac{K B (A-1)}{(A+1)^2} \frac{1}{a^{\frac{3(A+1)}{2}}} ,  \; \; 
p_3 = \frac{AK^2}{(A+1)^2} \frac{1}{a^{3(A+1)}}.
\end{equation}
Comparing with the barotropic EoS given by $p_i = \omega_i \rho_i$ one obtains $\omega_1= -1$, $\omega_2= \frac{A-1}{2}$ and $\omega_3= A$.
 Thus, the parameter $A$ plays an important role in determining the composition of the fluids in the universe. For example, $A=\frac{1}{3}$ leads to a universe with radiation, exotic matter and dark energy, $A=0$ leads to dark energy, exotic matter and dust. Thus once the EoS parameter $A$ is fixed the composition of the fluid in the universe gets determined. In order to obtain a viable scenario of the universe we consider interacting fluids model in the next section so that a transformation of one kind fluid in the later epoch gives rise to the composition of matter that we observe today.

\section{Cosmological Models}

In this section we consider two different models of interacting fluids in an
 EU scenario.

\vspace{0.5 cm}

{\centering {\bf Model I : The two fluids model}}
\vspace{0.5 cm}

In this case, we consider two interacting fluids with densities $\rho$ and $\rho'$ respectively which can exchange energies 
with each other. One of the fluid with energy density, say $\rho$ is dominated to begin with satisfying  a non-linear EoS given by eq. (1) which leads to an
emergent universe model as discussed above with no interaction. The contribution of the other fluid in the energy density of the universe  is assumed to be important at a later epoch.
The corresponding pressure of the former fluid is  given by
\begin{equation}
\label{eos1}
p= A \rho - B \rho^{1/2}.
\end{equation}
where $A$ and $B$ are constants. The other fluid satisfies a barotropic equation of state which is given by
\begin{equation}
\label{e1}
p' = \omega' \rho'
\end{equation}
where $\omega'$ corresponds to EoS parameter. The Hubble parameter $\left(H= \frac{\dot{a}}{a}\right)$ evolves according to the Friedmann equation which is given by
\begin{equation}
\label{f2}
3 H^2 =  \rho + \rho'.
\end{equation}
 In this  case we consider a cosmological model where exchange of energy between two different fluids is allowed.The are many astrophysical and cosmological motivations for considering
energy exchanges between the various components of the universe [39-44].
Different phenomenological considerations dictate the onset of such
interactions. For example, in various scalar field models of dark
energy, such as in quintessence, or k-essence, there arise phase transitions
during particular eras resulting in decay of the cosmological vacuum
energy, as well as particle production. Similarly, there could be other
cases of energy exchange, such as due to the evolution of a population of
primordial black holes whose evaporation time depends on the particular
formation mechanism or formation era. In the present paper, without
assuming any specific mechanism for energy exchange, we assume that the
interaction starts at some particular time  $t_i$.
 The two interacting fluids respect a total energy conservation equation and their densities evolve with time as 
\begin{equation}
\label{f3}
\dot{\rho} + 3 H (\rho + p) = - \alpha \rho H,
\end{equation}
\begin{equation}
\label{f4}
\dot{\rho'} + 3 H (\rho' + p') =  \alpha \rho H
\end{equation}
where $\alpha$ represents a coupling parametrizing the energy exchange between the fluids. One may view the above interaction as a flow of energy from first 
kind of fluid to the second one (say, dark matter) beginning at the epoch considered here. Now, using eq. (14) in eq. (17), we get a first order differential equation which can be integrated to obtain the behaviour of energy density in terms of the scale factor of the universe and the interaction coupling factor.  Thus the energy 
density and pressure  for the fluid of the first kind are given by 
\begin{equation}
\rho = \frac{B^2}{(A+1+ \frac{\alpha}{3})^2} + \frac{2 K B}{(A+1+\frac{\alpha}{3})^2}  \frac{1}{a^{\frac{3(A+1+\frac{\alpha}{3})}{2}}} + \frac{K^2}{(A+1+\frac{\alpha}{3})^2} \frac{1}{a^{3(A+1+\frac{\alpha}{3})}}, 
\end{equation}
\begin{equation}
p  = - \frac{B^2}{(A+1+\frac{\alpha}{3})^2} + \frac{K B (A-1+\frac{\alpha}{3})}{(A+1+\frac{\alpha}{3})^2}  \frac{1}{a^{\frac{3(A+1+\frac{\alpha}{3})}{2}}} + \frac{(A+\frac{\alpha}{3}) K^2}{(A+1+\frac{\alpha}{3})^2} \frac{1}{a^{3(A+1+\frac{\alpha}{3})}}.
\end{equation}
If the interaction is with a pressureless dark fluid {\it i.e.,} $p'=0$ (however, $\rho' \neq 0$), eq. (18) can be integrated using eqs. (16) and (19) which determine the total energy density and pressure as follows : 
\begin{equation}
\rho_{total} = \rho + \rho' = \frac{B^2}{A+1} + \frac{2 K B}{(A+1)^2}  \frac{1}{a^{\frac{3(A+1)}{2}}} + \frac{K^2}{A+1} \frac{1}{a^{3(A+1)}}, 
\end{equation}
\begin{equation}
p_{total}  = p  = - \frac{B^2}{A+1} + \frac{K B (A-1)}{(A+1)^2} \frac{1}{a^{\frac{3(A+1)}{2}}} + \frac{A K^2}{(A+1)^2} \frac{1}{a^{3(A+1)}}.
\end{equation}
The equation of state parameter for the second fluid is given by 
\begin{equation}
\omega'=\frac{p'}{\rho'} = \frac{p_{total}-p}{\rho_{total}-\rho}.
\end{equation}
In the limiting case as $\omega' \rightarrow 0$, we get $p_{total} = p$.  
An interesting case emerges when the coupling parameter $\alpha = 2$ and $A= \frac{1}{3}$. In this case a universe with dark energy, exotic matter and radiation to begin with (i.e., before the interaction sets in) made a transition to a matter dominated phase. Thus,  an emergent universe with radiation, dark energy (component that is subdominant at
early times) and exotic matter (for $A=\frac{1}{3}$) to begin with transits  to
 a universe with matter domination phase after an epoch after $t > t_i$ for the interaction coupling strength $\alpha = 2$. Hence, a consistent scenario of the observed universe in the EU model may be realized in this case.

\vspace{0.5 cm}

{\centering {\bf Model II : The three fluids model}}
\vspace{0.5 cm}

The original EU model was obtained in the presence of non-interacting fluids permitted by the parameter $A$ in a flat universe case.  The corresponding densities and pressures are given by eqs. (\ref {r1}) and (\ref {r2}) respectively.  For
non-interacting fluids, the EoS parameters for the three fluids permitted above are given by
$
\omega_1=  -1, \; \; \omega_2=  \frac{1}{2} \left( A -1 \right), \; \;  \omega_3=  A. $   For $0 \leq A \leq 1$, it accommodates dark energy, exotic matter and 
the usual barotropic fluid. The energy density and pressure of the exotic matter and that of the barotropic fluids decreases with the expansion of the universe. However, the rate of decrease is different evident  from eqs. (\ref {r1}) and ( \ref {r2}). We assume an interaction among the components of the fluid   in the universe which is assumed to be originated at a later epoch (such interactions could arise due to a variety of mechanisms \cite{bar,cim,jam,lip,cost,cot}). Assuming onset of interaction among the composition of the fluid at $t \geq t_o$, the conservation equations for the energy densities of the fluids  now can be written as
\begin{equation}
\label{f8}
\dot{\rho_1} + 3 H (\rho_1 + p_1) = - Q',
\end{equation}
\begin{equation}
\label{f9}
\dot{\rho_2} + 3 H (\rho_2 + p_2) =  Q,
\end{equation}
\begin{equation}
\label{f10}
\dot{\rho_3} + 3 H (\rho_3 + p_3) =  Q'-Q,
\end{equation}
where $Q $ and $Q'$ represent the interaction terms, which can have arbitrary 
form, $\rho_1$ represents dark energy density, $\rho_2$ represents exotic matter and $\rho_3$ represents normal matter. In this case $Q < 0$ corresponds to energy transfer from exotic matter sector  to two other constituents, $Q' > 0$ corresponds to energy transfer from dark energy sector to the other two fluids, and $Q' < Q$  corresponds to energy loss for the normal matter sector. The case $Q=Q'$ corresponds to the limiting case where dark energy interacts only with the exotic matter.  It is important to see that although the three equations are different the total energy of the fluid satisfies the conservation equation together. It is possible to construct the equivalent  effective uncoupled model, described by the following conservation equations:
\begin{equation}
\label{f11}
\dot{\rho_1} + 3 H (1+ \omega_1^{eff} ) \rho_1  = 0
\end{equation}
\begin{equation}
\label{f12}
\dot{\rho_2} + 3 H (1+ \omega_2^{eff} ) \rho_2  =  0
\end{equation}
\begin{equation}
\label{f13}
\dot{\rho_3} + 3 H (1+ \omega_3^{eff} ) \rho_3  = 0
\end{equation}
where the effective equation of state parameters are given below:
\begin{equation}
\label{f14}
\omega_1^{eff} = \omega_1 + \frac{Q'}{3 H \rho_1},
\end{equation}
\begin{equation}
\label{f15}
\omega_2^{eff} = \omega_2 -  \frac{Q'}{3 H \rho_2},
\end{equation}
\begin{equation}
\label{f15}
\omega_3^{eff} = \omega_3 + \frac{Q - Q'}{3 H \rho_3}.
\end{equation}
Now, if we consider the interaction as $Q- Q'= - \beta H \rho_3$,  the 
effective state parameter for the normal fluid becomes
\begin{equation}
\omega_3^{eff} = \omega_3 - \frac{\beta}{3}
\end {equation}
In fig. 1 we plot the variation of effective equation of state 
parameter $\omega_3^{eff}$
with $\omega_3$  (which corresponds to $A$  of the EoS parameter) for different  strengths of interaction determined by $\beta$. We note that  as the strength of interaction is increased the value of $\omega_3$  ({\it i.e.,  $A$})  for which $\omega^{eff}=0$ (corresponds to matter domination) is found to increase. Thus a universe with any $A$ value is found to admit a matter dominated phase at a late epoch depending on the strength of the interaction which was not permitted   in the absence of interaction in an EU model proposed by Mukherjee {\it et al.} \cite{euorg}. 
It may be pointed out here that in the very early era a universe is assumed with a composition of three different fluids having no interaction in this picture, thus the behaviour of  the universe at  early times remains unchanged as was found in the original EU model.  Thus the emergent universe scenario proposed by Mukherjee {\it et al.} \cite{euorg}  can be realized in the early era but  at a later epoch the composition of matter changes  in the present scenario  from  its original one with the onset of interaction. 
This feature represents a clear improvement over the earlier cosmological scenario in an emergent 
universe \cite{euorg,eust,eugb} where it is
rather difficult to accommodate a pressureless  fluid.

\begin{figure}
\begin{center}
\includegraphics{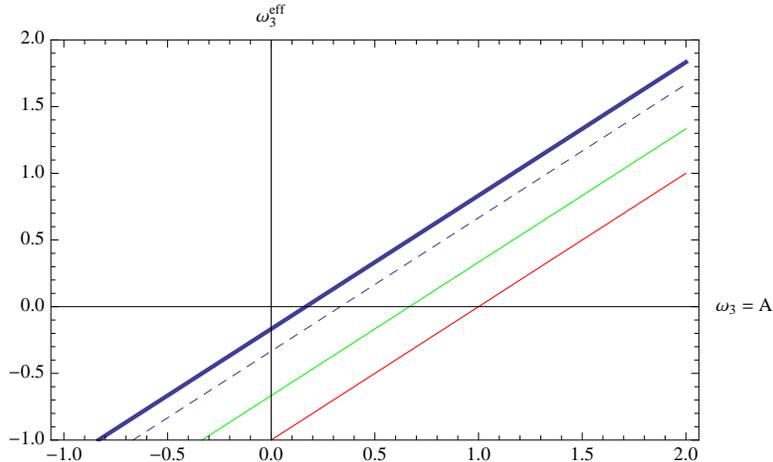}
\caption{Plot of $\omega_3^{eff}$ with EoS parameter $A$ for different interaction $\beta$.  The thick, dash, red line and green lines are for $\beta=0.5, \; 1, \; 2, \; 3$ respectively.}
\end{center}
\end{figure}



\section{Generalized Second Law of Thermodynamics in Emergent Universe Model}

In the above analysis emergent universe models have been discussed in the presence of different types of interaction among the fluids. In the present section we investigate consistency of thermodynamic properties considering the universe as a thermodynamical system. For a flat geometry the apparent horizon coincides with the Hubble horizon which is given by
\begin{equation}
 r_A= \frac{1}{H}
\end{equation}
In general, the apparent horizon is a function of time. Thus, a change in the
apparent horizon leads to a change in volume, and consequently, the energy and entropy will change by $dE$ and $dS$ respectively. The energy momentum tensor before and after the change are described by the same $T_{\mu \nu}$, and we can consider that the pressure and the temperature remains the same \cite{fro,fro1,fro2,fro3,fro4}. The first law of thermodynamics for the fluids considered here are given by
\[
 dS_1=\frac{1}{T} \left(p_1 dV+ dE_1\right)
\]
\[ 
 dS_2=\frac{1}{T} \left(p_2 dV+ dE_2\right) 
\]
\begin{equation}
 dS_3=\frac{1}{T} \left(p_3 dV+ dE_3\right)
\end{equation}
where the volume of the system $V= \frac{4 \pi r_A^3}{3}$ is bounded by the 
apparent horizon, and thus $dV= 4 \pi r_A^2 dr_A$.
Hence, the rate of change of entropy for the above fluids are given by
\begin{equation}
 \dot{S}_{1} = \frac{1}{T} \left(4 \pi r_A^2 \dot{r_A} p_1 + \dot{E_1}\right), \;\;
\dot{S}_{2} = \frac{1}{T} \left(4 \pi r_A^2 \dot{r_A} p_2 + \dot{E_2} \right), \; \;
\dot{S}_{3} = \frac{1}{T} \left(4 \pi r_A^2 \dot{r_A} p_3 + \dot{E_3} \right) 
\end{equation}
where 
\begin{equation}
 \dot{r_A} = \frac{1}{2} r^2_A \left( (1+\omega_1) \rho_1+ (1+\omega_2) \rho_2 +(1+\omega_3) \rho_3 \right)
\end{equation}
which is obtained differentiating the equation
\begin{equation}
 \frac{1}{r_A} = \frac{1}{3} \left( \rho_1+\rho_2+\rho_3 \right)
\end{equation}
and then making use of the conservation equations given by eqs. (27)-(29) with  eq. (34).
The corresponding energy density and pressure are given by
\[
 E_1= \frac{4 \pi}{3} r^3_A \rho_1
\]
\[
E_2= \frac{4 \pi}{3} r^3_A \rho_2,
\]
\begin{equation}
E_3= \frac{4 \pi}{3} r^3_A \rho_3 
\end{equation}
and
\begin{equation}
 p_1= \omega^{eff}_1 \rho_1,  \; \; \; p_2= \omega^{eff}_2 \rho_2,  \; \; \; p_3= \omega^{eff}_3 \rho_3 .
\end{equation}
Using the time derivative of the apparent horizon we get
\[
 \dot{S}_{1} = \frac{4 \pi r^2_A}{T}  \rho_1 (\dot{r}_A - H r_A) (1+ \omega^{eff}_1)
\]
\[
\dot{S}_{2} = \frac{4 \pi r^2_A}{T}  \rho_2 (\dot{r}_A - H r_A) (1+ \omega^{eff}_2,
\]
\begin{equation}
\dot{S}_{3} = \frac{4 \pi r^2_A}{T}  \rho_3 (\dot{r}_A - H r_A) (1+ \omega^{eff}_3).
\end{equation}
According to the generalization of black hole thermodynamics \cite{fro3,fro4,gwg}   to a cosmological framework, the temperature of the horizon  is related to its radius \cite{jac,jac1,jac2 }  as
\begin{equation}
 T_h= \frac{1}{2 \pi r_A},
\end{equation}
leading to the rate of change of entropy given by 
\begin{equation}
 \dot{S}_h= 16 \pi^2 r_A \dot{r_A}.
\end{equation}
It may be mentioned here that the Hawking radiation of apparent horizon in  a FRW universe is computed by Cai {\it et al.} \cite{cai}.
The rate of change of total entropy becomes
$ \dot{S}_{total} = \dot{S}_1+ \dot{S}_2+ \dot{S}_3+ \dot{S}_h$,
which leads to
\begin{equation}
 \dot{S}_{total} = 4 \pi^2 r^6_A H \left[ (1+ \omega_1)\rho_1 + (1+ \omega_2)\rho_2 + (1+ \omega_3)\rho_3 \right]^2 \geq 0. 
\end{equation}
The non-negativity of the time rate of change of $S_{total}$ demonstrates 
the validity of the second law of thermodynamics in the context of the emergent universe model.

\section{Discussion}
In this paper we have investigated cosmology of the emergent universe scenario 
in the presence of 
interacting fluids. The purpose of the present analysis is to demonstrate
the possibility of obtaining viable cosmological dynamics of the emergent
universe.
Two different cosmological models have been presented here. In 
Model I, we consider the flow of energy from  the fluids required to realize 
the emergent universe to a pressureless fluid which sets in at an 
epoch $t=t_i$. The density of the pressureless fluid assumes importance 
as matter component after the epoch $t_i$. In Model II, we consider 
interactions among the three fluids of the emergent universe at time $t=t_0$. 
Before this epoch the emergent universe can be realized without an interaction 
among the fluids. The problem with earlier cosmological realizations of the
emergent universe was that once the EoS parameter
 $A$ is fixed at a given value, the universe is unable to come out of the phase 
with a given composition of fluids. In the present work we overcome this problem
by assigning an interaction among the fluids at the epoch $t_0$.   A 
cosmological evolution of the observed universe through unified dynamics of
associated matter and dark energy components thus becomes feasible in the
emergent universe scenario. In fig. (1), we plot variation of $\omega_3^{eff}$ with EoS parameter $\omega_3=A$ for different interaction. It is evident that a early universe with a radiation dominated phase transits to a matter dominated phase with all the features observable at the present moment with the onset of interaction considered here.
Consistency of the interacting 
fluid emergent universe scenario with the generalized 
second law of thermodynamics is also shown. Further work is needed in order
for a comparative analysis of the interacting fluid emergent universe cosmology
with more popular current cosmological models. Detailed analysis of 
observational constraints pertaining to various eras of the universe are
expected to yield bounds on the parameters of the emergent universe models 
considered here, leading to a firmer assessment of the viability of such models.

\section{Acknowledgement:}

BCP would like to thank S. N. Bose National Centre for Basic Sciences, Kolkata for warm 
hospitality during a visit when the work was initiated.  BCP likes to thank Institute of Mathematical Sciences (IMSc), Chennai for hospitality during a visit. BCP also acknowledges University Grants Commission (UGC), New Delhi  for  financial support awarding with a Major Research Project (Grant No. F. No. 42-783/2013(SR),  dated 13 Feb 2014). 

\pagebreak

\end{document}